# Gnutella: Topology Dynamics On Phase Space

Chunxi Li and Changjia Chen
School of Electronics and Information Engineering, Beijing Jiaotong University, 3 Shangyuancun, Haidian
District, Beijing 100044, P. R. China
cxl@telecom.njtu.edu.cn and changjiachen@ sina.com).

*Abstract*—**In this paper, the topology dynamic of Gnutella is studied through phase space. The dynamic changes in peer degree are studied as a time series in two dimensional phase space which is defined as the number of connected leaves and the number of connected ultras. The reported degrees are concentrated in three special Software related regions that we named as Ultra Stable Region, Leaf Stable Region and Transition Belt. A method is proposed to classify degree traces in phase space into different classes. Connection churn then is studied along with the churn in degree. It shows that the topological structure of Gnutella is rather stable in its connection degree but not the topology itself. The connection drop rate is estimated and the live time of connections is inferred afterwards. M/M/m/m loss queue system is introduced to model the degree keeping process in Gnutella. This model revealed that the degree stable is ensured by large new connection efforts. In other words the stable in topological structure of Gnutella is a results of essential unstable in its topology. That opens a challenge to the basic design philosophy of this network.**

## I. INTRODUCTION

DURING recent years, the increasing popularity of peer to-peer (p2p) networks has led to growing interest in measurement based characterization of popular p2p systems (*e.g.*, [1~6]). These characterizations provide deeper insight into the behavior of p2pP systems, essential for proper design and effective evaluation. These characterizations become more important as p2p systems rapidly increase in use over the Internet.

Gnutella as one of the largest p2p networks represents a special p2p network design philosophy of unstructured construction. In this philosophy, the network topology is basically unspecified except several very basic structural parameters such as two layers and guidance in connection degrees. To many people's surprise, such a simply designed network has accommodated tens of millions peers and stably operated for at least half decade. It's success in stability and popularity is certainly not only the enthusiastic in free content share, some its topological characters contribute this success even though these characters are not known very well and the topology of this network itself is almost not designed. In other side, content pollution and p2p DDOS has been a new topics raised many concerns in recent years [11-13]. Studies show that no matter pollution or antipollution, attack or defense, topology is always one of most relevant character must be taken into considerations, Gnutella is often one of such network stands for a special class of unstructured topology.

A new measurement philosophy and analysis methodology to Gnutella network is introduced in this paper. The contributions we made in this paper can be summaries as:

***Continue Watch vs. Snapshots:*** Snapshot taken is a common technique [8~10] to characterize a p2p overlay. The capture the overlay topology is a graph, with peers as vertices and connections as edges. The topology characters can be analyzed by examining the measured snapshots. Even though there are many literatures on the topology of Gnutella, but the structure of this topology is still confused in our knowledge. For example, the degree distribution as a very basic aspect of topology has been reported in at leas three different versions. All of them are based on certain real measurements. Connection churn is notice by almost all studies on Gnutella topology. But how does and why is this churn has not been answered clearly. We believe that snapshots only are not enough to understand the dynamics of Gnutella topology. Different measurement scenarios are needed to measure the network more consistently and insistently beyond snapshots taken.

In this paper we will report one of such efforts in our Lab. The measurement is taken about 23hours, and about 10,000 peers are selected from our scope crawling conducted before. The selection is somehow randomly balanced on peers software version, mode and reported degrees. The selected peers are crawled periodically at a time interval of 30 minutes. Only 5,492 peers are ever responded in our crawl. Our study on Gnutella topology in this paper is based on this measurement result.

***Phase Space vs. Peer Degree:*** It is well known that the peer connection degree is an important parameter to the topology of Gnutella. But different findings were reported at literatures on this topic. Instead of simply use the number of connected peers to define the topological degree, in this paper, a phase space is introduced to divide the topological degree as the number of connections to leaf peers (the leaf degree or simply Ldeg) and the number of connections to ultra peers (the ultra degree or

This work was supported by the Chinese NSF under Grant 60132030 and 60202001.

Authors are with School of Electronics and Information Engineering, Beijing Jiaotong University, 3 Shangyuancun, Haidian District, Beijing 100044, P. R. China  (phone: 86-10-51891873; e-mail: cxl@telecom.njtu.edu.cn and changjiachen@ sina.com).

simply Udeg) since the protocol of Gnutella documents them separately with different connection rules. A state or a point in the phase space is a pair of integers $(d_l, d_u)$. What we say that, a peer has a state $s_i=(d_l, d_u)$ or at the point of $p_i=(d_l, d_u)$ in phase space at a given time $t$, means that the peer connects to $d_l$ leaf peers and $d_u$ ultra peers at this time. By this way, topological evolution of a peer can be described as a time-state trace in this phase space. The characters of topological degree can be explained through protocol specifications directly.

*Interpret Protocol Features of Gnutella Topology:* The whole topology of Gnutella network is generated and kept by peers individually. A set of simple rules is specified in protocols to guide peer how to connect and admit connections to other peers [18]. Maximum number of connections to leaf peers and to ultra peers is specified separately for leaf mode and ultra mode respectively. If the connected number of peers reaches this number, a peer will never admit new connection. We will denote this number as $U_l$ and $U_u$ for leaf peer and ultra peer connections respectively for an ultra mode peer. Similarly, $L_l$ and $L_u$ for leaf peer and ultra peer connections respectively for a leaf mode peer. For ultra mode peer, it always waits for other leaf peers to initiate the leaf connections. But there is a number $D_u$. If the connected ultra peers are less than $D_u$, a peer will try to find and connect to other ultra peers by himself based on a previously recorded peer list. If the connected ultra peers beyond $D_u$, this peer will passively wait other peers to initial the connection. Up to our knowledge, except peer degree distribution, these protocol features have never been measured and analyzed before. Through peers distribution on phase space we found there is two stable points around $(U_l, U_u)$ and $(L_l, L_u)$ corresponding to the leaf and ultra modes of a peer may play in the network respectively. A relatively narrow transition belt in the phase space can be observed clearly, which show the path that a peer up to ultra from leaf or drop to leaf from ultra. These two stable points and the transition belt is slightly different for different software implementations such as LimeWire and BearShare. Hence we believe that peer distribution on phase space will be a useful tool in analysis the topological structure of Gnutella along with its protocol specifications. The transient character of peer distribution on phase space is studied in this paper. A partitioning rule to the phase space is introduced in this paper as well.

*State Transient, Peer Traces vs. Degree Distribution:* Peer distribution can be drawn from ether continues watch or snapshot. But only continues watch can give out the degree evolution of each individual peer and the state transient probability as a whole. Since there is no any analytical result to ensure the stationary evolution of Gnutella topology, we don't know if the peer distribution based on one snapshot, the peer distribution based on continues watch, and a degree trace of individual peer will be looks alike or not with a large probability. In this paper we use examples to show the similarity on the peer distribution based on one snapshot and based on continues watch. We also discussed the presentation and classification of peer traces. We find most peers' degree traces are stable with occasional drift when they stay in one mode. Mode switch is a relatively fast process. Peers rapidly go from leaf nodes to transition belt being peered with many ultra peers, but take a while to find enough leaves. Even though our work in this topic is far from complete, but based on these preliminary observations, we believe that the topology evolution of Gnutella is relatively a stationary process.

*Topology Structure vs. Topology:* In most literatures, people use degree distribution and other static statistics (e.g. a list at table 3 in [19]) to describe the topology of the studied network. Essentially all these statistics describe the structure of topologies for a class of graphs not the topology itself. It is fine for slow changed network such as AS level Internet but unnecessarily true for Gnutella such a fast churned network. An important issue but ignored by many literatures is the degree and the actual connections, more precisely, the degree churn and connection churn. Our measurement shows that, the state trace of a Gnutella peer is looked like a controlled object. In most cases, the courses they followed are roughly same. A small disturb will be quickly corrected. In other side, the actual connections are changed more intensively and randomly in general. Our measurement tells us that the connection churn is much more easily observed between two consecutive crawls than that of degree churn, and the number of changed connections is much bigger than the number of changed degrees. In this sense the Gnutella is stable in its topological structure (such as *the degree for each peers, clustering, distance, etc.* many static features of a graph for example listed at [19] at Table 3) but not in its topology (the *edges of peer to peer connection*). The appearance of the topology looks alike but the detailed connections might be totally different at different observation time. As an interesting example, in our measurement, 5,492 peers are continuously crawled 47 times during 23 hours, within them 147,094 queries are responded with 5,874,495 unique peers reported. In average 39.937 unique peers reported per query. In opposite, in the half an hour snapshot (trace *crawl_2005_10_13_11_05*) taken by [8-10], more than 700K peers are queried and each peer only queried once. About 317,604 queries are responded with 14,201,440 unique peers reported. In average 44.7143 unique peers reported per query. As a graph, the 47 snapshots in our measurement and the snapshot taken by [8-11] are all looks alike up to static statistics. But the number of peers ever connected to a specified peer in two different snapshots, are almost the same as the number of peers connected to two different peers within one snapshot. This phenomenon will significantly impact to the file distribution [20] and file query in Gnutella. We will study it in our further work. In this paper, we will compare the degree churn and connection churn, infer the connection's live time and peer's live time, and try to model the degree keeping process.

*Multiple Serves Queue Modeling to The Degree Keeping Process:* In software implementation, each connection is managed as a channel slot. The number of channel slots for leaf peers $U_l$ ($L_l$) and for ultra peers $U_u$ ($L_u$) is a parameter previously sited. The factory default value for LimeWire is $U_l$=30 and $U_u$=32 for ultra mode. In most cases an ultra mode peer will passively waiting to other peers to initiate a connection. If there is channel slots left the peer will admit this connection otherwise the applied connection will be rejected. Once a connection is dropped, the channel slot it occupied before will be counted as an empty slot and can be used to admit new connection application again. Take the terminology in queuing theory, each channel slot can be thought as a server, connection application can be thought as customers arrive, and the connection live time can be thought as service time of a customer. The degree keeping process can be modeled as a multiple servers queue system in principle. The analysis in this paper will show that the connection live time is roughly exponentially distributed and connection application is roughly Poisson. Hence we think a memory less M/M/m/m loss queue can be a simple model to approximate the degree keeping process. Even though there is certain evidence to doubt the precision of this model, but we still think it is a good model due to its simplicity and due to there is no any existent analytical model yet up to our knowledge. By our judgment, the challenge to M/M/m/m model does not come from the multiple server queue model but the memory less nature in the system. First, certain measurements reported qualitatively that the connection live time is some how related to the live time of peers made this connection [9]. The longer of peers live the longer the connection live time between them. Our data shows most connections are short lived compared with peers' live. Hence we believe the correlation in connections live and peers' live will not hurt the precision of this model very much. Second, the leaf degree and ultra degree are not independent process. To use two independent M/M/m/m systems to model leaf and ultra degree keeping process respectively will miss the dependency between these two processes. We will study this in our future works. Last, the degree keeping process for ultra peer connections is far more complex than leaf peer connections in software implementations. At least when the number of connected ultra peers below the threshold $D_u$, a peer will connect to other ultra peers actively. The queue model is not fitted in this case. This will hurt our queue model mostly. Our result is also show that, the M/M/m/m loss queue model fits leaf process much better than ultra process. Relatively big differences exist at the degrees near $D_u$ for ultra peer process between the measured result and the calculated result from the model. This is a study work we are taking now to improve the model especially for ultra degree keeping process.

The most important insight from our M/M/m/m model is that, the controlled nature in degree is due to large demands for new connections. If the demands dropped into certain threshold, the degree cannot be stabled. Especially for ultra peer connections, a dropped ultra peer connection can only be compensated by ultra peers with a ultra degree less than $D_u$. Most of them will be newly upped peers from leaf mode. That means the degree stability around ($U_l$, $U_u$) might imply the instability that needs large amount of leaf peers up to ultra peers.

***Optimal linear prediction***: Optimal linear prediction then is discussed in this paper following standard statistical filtering theory.

The key findings of this paper include: The use of phase space allows identifying the stable (a leaf peer stays as leaf and an ultra peers stays as ultra) and transition states of the Gnutella topology well. Most peers stay connected using the default settings present in the software and this dictates the topology more than any other factor. The lifetime of a connection is much shorter than the lifetime of the peer.

## II. II. PHASE SPACE OF GNUTELLA TOPOLOGY

### 2.1 Phase Space and Phase Space Representation

Figure 1 depicts the measurement result of trace *topo-0215 06.9.30-021606.8.15*, which is taken on Feb. 15-16, 2006 and last 23 hours. In this measurement, 5,492 peers are crawled every half hour and with 147,094 responses in total (in average 26.8 answers per peer, it is about 1 answer/1 hour roughly for each peer). Distribution of Ldeg and Udeg is shown in fig 1(a). The phase space presentation is in fig 1(b). Fig 1(c) and (d) use different visualization method to describe the intensive in the phase space. To emphasis small intensive, the height in fig 1(c) is the $4^{th}$ root of the actual value. In fig 1(b), different (color, mark) combinations are used to indicate the mode (leaf or ultra) of a peer at the time the state is reported. Since not all answered peers told us their mode, a large amount of (green +) points in this sub graph indicate the unknown mode states. It can be seen clearly that, the reported states are concentrated at several points and the peer modes at each given points are almost the same if it is known. So we can infer the mode of unknown mode states from their position in phase space. The first advantage to use phase space description comes from this observation that peer mode is strongly correlated to its phase state at any observation time.

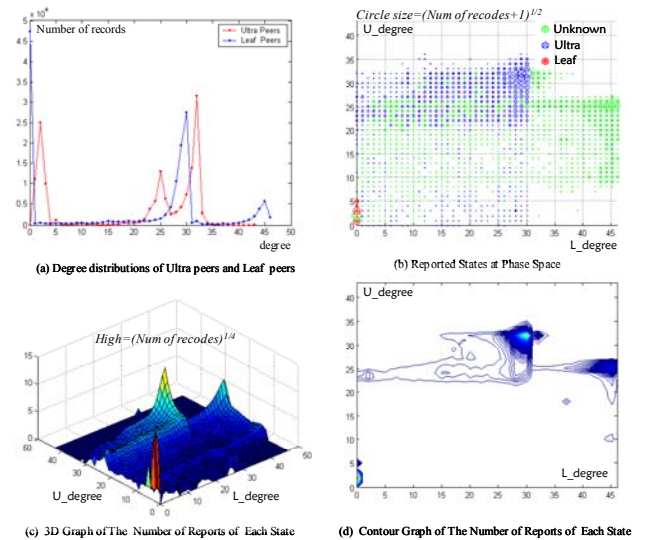

(a) Degree distributions of Ultra peers and Leaf peers  
(b) Reported States at Phase Space  
(c) 3D Graph of The Number of Reports of Each State  
(d) Contour Graph of The Number of Reports of Each State  

Fig 1. Presentation Measurement Resells On Phase Space

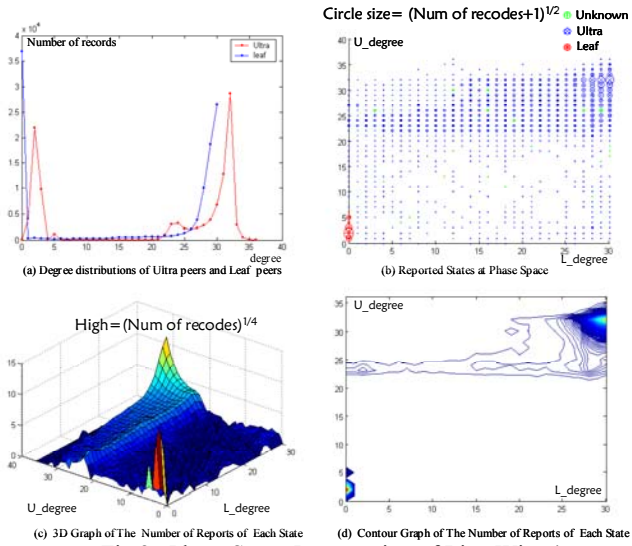

Fig 2. Phase Space Presentation of LimeWire 4.x

Looking at figure 1 carefully we can find that, there seems two major models here. In fact there are 3,437 LimeWire 4.x peers with 109,480 records (31.9 answers/ peer) and 1,845 BearShare peers with 31,008 records (16.8 answers/peer) in our measurement data set. The number of peers and records of LimeWire 4.x and BearShare together has been 96.2% and 95.5% of the total peers and records respectively. The topology character is mainly affected by the behavior of these two software versions.

**2.2 Stable Points and Transition Belt:** Figure 2 shows the phase space presentation for LimeWire 4.x in our trace *topo-021506.9.30-021606.8.15*. From this figure we can see that there are two concentrated regions at upright and bottom left corners, and a horizontal belt at up plane. The centers of the concentrated regions are at (0,2) and (30,32) for LimeWire *( at (0,1) and (45,25) for BearShare the figure is omitted since space limitation)*. We will name these two points as leaf stable point (bottom left) and ultra stable point (upright) respectively and the belt as transition belt. Stable points and transition belt are software related and rather consistent at different measurements. The transition belt at phase space is about the region $23 \leq d_u \leq 32$ for LimeWire *( and $23 \leq d_u \leq 27$ for BearShare)*. Transition belt is the main channel peer experienced when they change modes from leaf to ultra or form ultra to leaf. If look at the transition belt more carefully, we can find a triangle region at up left corner. Peers from leaf up to ultra will mainly step to these region firs. For LimeWire, the transition belt is consisted by two parts: upper belt and lower belt. The lower belt is the peer up channel and the upper belt is mainly due to churn of ultra peers at stable point. There is also an outstanding point at $d_l$=1~3 and $d_u$=25 for LimeWire. We sense this is due to the design criterion adopted in this software. The peer up to ultra from leaf will quickly reach this point by connecting to ultra peers it has found before and then stay in this belt to wait other ultra peers to connect him. The process of connecting others is much fast than that of waiting to be connected. Then transient into ultra stable point along lower belt will take longer time than up to the belt from leaf state. We will call this point as *up point*.

Different measurement philosophy is taken in [8-10,14], where massive peers are crawled within very short of time interval instead of small number of peers is repeatedly crawled in a long time in our measurement. The phase space presentation for such a snapshot (trace *crawl_2005 _10_13_11_05* measured by [8-10,14]) is similar to figure 2 and omitted here since space limitation. It crawled about 350K peers (mostly ultra) within half an hour. That means the stable points and transition belt are protocol features independent to the measurement scenarios. That also means that the degree process in Gnutella is relatively stationary such that the time statistics (by our measurement) are similar to the assemble statistics (by snapshots).

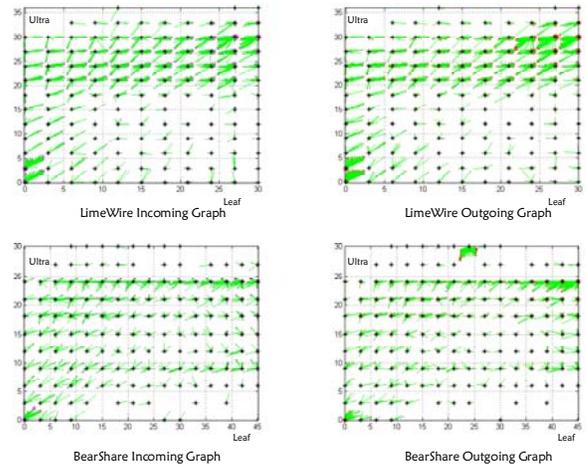

Fig 3. Incoming and Outgoing Graph

**2.3 State Transient:** Figure 1 and 2 are only static descriptions of the Gnutella topology. Recorded (*starting state*, *ending state*) pairs can be used to describe topology dynamics of Gnutella peers. We use incoming graph and outgoing graph in figure 3 to draw the recorded state transition behaviors respectively. In the incoming graph, every points in the regular lattice represent an ending state, and the lines into one point represent the direction and proportionally shirked distance of starting states that transient into this ending state. Oppositely, point in outgoing graph stands for starting state, and lines indicate the transition from give starting state to different ending states. In order to make the resulted graph not too messy, we grouped 2x2 states as one point in figure 3. From this graph we can see that, the state transition seems following certain rule, no matter at incoming graph or outgoing graph, there are mainly two transition directions and pointing to two stable points respectively. Exceptions are at the two stable points. The transition direction there is more diverse, but all of them point into transition belt or another stable point. One attribute unable to present by both of incoming and outgoing graph is the stationary probability of a state, which tell us how long a peer may stay in this state. The contour graph of stationary probability is depicted at figure 4 and compared with the contour graph of static intensive graph. It is interested to notice that, the stationary probability at *up point* is significant higher than that of other points except stable points. It indicates that, a peer is much more easily to find ultra peers to connect than find leaf peers during transition up into an ultra peer. And if the peer happen fall in the region of up point, it is much more harder to find peers to connect. A crowding at this points might be a indication of abnormal network condition that too many ultra peers generated at this moment compared with the number of leaf peers. A point at (37,18) for BearShare outstands at all of these three graphs on this software. The reason we still not know yet, but think might be related to churn in topology.

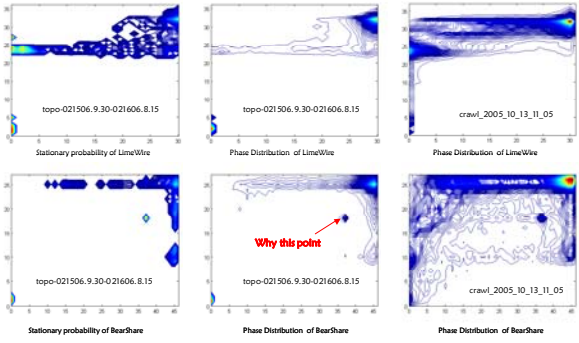

Fig 4. Stationary Probability and Compared with Intensity

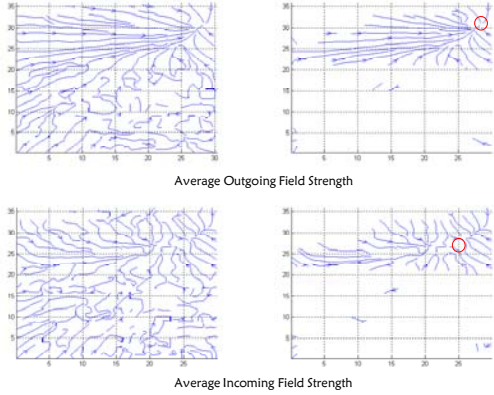

Fig 5 Average stream direction For LimeWire 4.x

Figure 5 takes a different visualization method to present the contents in figure 3. The average transition is interpreted as stream direction. The subplots on the left are based on all recorded data and the subplots on the right only plot those origin states at least 10 recorders on this state being observed. The transition belt and likely transition paths can be seen clearly on this figure. In the outgoing graph the ultra stable point is an obvious tractor. There is no obvious tractor at the incoming graph, but it seems that, certain force exists that draws peers away from stable point into transition belt.

**2.4 Partition On Phase Space:** Based on above observations, we can roughly partition phase space into 4 sub regions (figure 6): Leaf Stable Region (LSR), Ultra Stable Region (USR), Transition Belt (TB) and Ultra Degradation Region (UDR). In fact most peers stay at two stable regions. Peers in TB region are by two reasons: up to ultra from leaf or degradation from USR due to loss leaves. It can be observed frequently in measurements. Peers in UDR are rare and they are there due to loss ultras degradation from USR. We believe that to measure the intensive in these four regions will tell the current working condition of the Gnutella network. Uneven high counts in TB or in UDR indicate the disproportion in the number of leaf peers and ultra peers. A partition rule adopted in this paper is also marked in figure 10 to calculate transition statistics between these regions. For simplicity, we take each region as a rectangular and the boundary coordinates are marked in the figure. The number without brackets is for LimeWire and within the brackets is for BearShare. We will use the order of [*LSR, USR, TB, UDR*] to write mathematical formula. The resulted transfer probability $G$, the measured state distribution $p$ and the equilibrium distribution $h$ resulted from $G$ are listed below. The subscript $L$ and $B$ indicate it is for LimeWire or for BearShare respectively.

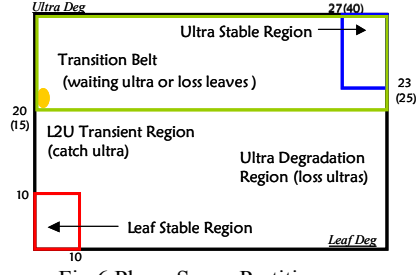

Fig 6 Phase Space Partition

$$G_L = \begin{bmatrix} 0.9878 & 0.0023 & 0.0401 & 0.0116 \\ 0.0029 & 0.9325 & 0.3666 & 0.0787 \\ 0.0089 & 0.0645 & 0.5880 & 0.1829 \\ 0.0005 & 0.0007 & 0.0052 & 0.7269 \end{bmatrix} \quad p_L = \begin{bmatrix} 0.3298 \\ 0.5305 \\ 0.1354 \\ 0.0042 \end{bmatrix} \quad h_L = \begin{bmatrix} 0.3955 \\ 0.5107 \\ 0.0901 \\ 0.0037 \end{bmatrix}$$

$$G_B = \begin{bmatrix} 0.9857 & 0.0025 & 0.0497 & 0.0104 \\ 0.0009 & 0.9616 & 0.3655 & 0.1062 \\ 0.0038 & 0.0196 & 0.5233 & 0.0916 \\ 0.0096 & 0.0163 & 0.0615 & 0.7919 \end{bmatrix} \quad p_B = \begin{bmatrix} 0.2994 \\ 0.5080 \\ 0.1152 \\ 0.0773 \end{bmatrix} \quad h_B = \begin{bmatrix} 0.2975 \\ 0.5901 \\ 0.0405 \\ 0.0720 \end{bmatrix}$$

**2.5 Summary of this section:** In this section we interpret our measurement result on phase space to argue that protocol features can be caught more easily and more clearly in this way, since the highlight of protocol events is specified by Ldeg and Udeg jointly. The protocol features we have identified in this paper are divided into static features and dynamic features. The static features include the two stable points and transition belt. Measurement shows that the practical Gnutella network is working at its default topological structure boundary. In other words, most peers in the network have fully connected with other peers constrained mainly by the default settings on protocol. State transient probability is the main dynamic feature discussed in this paper. Transient graph and stream graph on phase space are introduced to visualize the transient property instead of a numerical matrix to make it easier to feel. Dynamic feature shows that the transition belt is indeed the channel of peers up to ultra mode from leaf mode and the region of ultra mode peer drifting. Finally, a partition on phase space is suggested to divide the protocol statue of peers into 4 different classes. The transfer probability of these regions is calculated from the real measured data. The measured probability and calculated equilibrium probability based on transfer probability are compared as well. It evinces certain degree of stationary in the network dynamic. All above mentioned methodologies and results are new up to our knowledge. We believe it is useful in understanding the working statue of real Gnutella network. It also serves a good start point for further modeling and analytical studying the detailed behaviors of Gnutella.

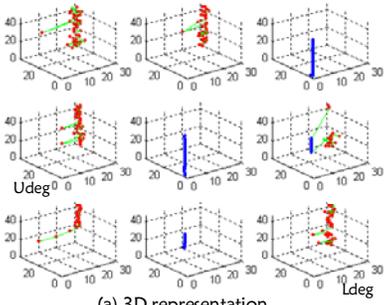
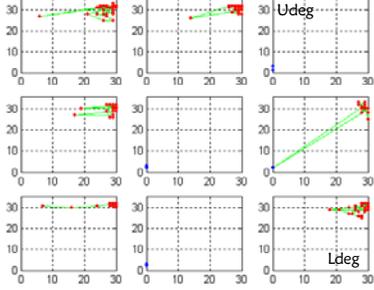

Fig 7 Peer Trace
(a) 3D representation
(b) 2D representation

## III. PEER TRACE

**3.1 Peer Trace:** For each peer we queried, we have a sequence of data, which records the peer's positions in phase space at each response time. Peer trace is defined as a piecewise connected line of this recorded sequence for each peers. No one has studied peer trace before. We study it because we think it is important in understanding the topology dynamic of Gnutella. Nowadays studies on Gnutella topology are based on assemble average. Most of them are concentrated on the degree distributions. A natural question to ask is whether the distribution is resulted by different kind of peers wanders at different degree regions or peers with similar property randomly drift within whole range. It matters in explain the protocol functions and the working mechanisms of network. This question can only be clear answer by peer trace studying. Our trace is too short to study the problem of state distributions on time and on assemble directly. Instead peer traces classification is studied in this paper. Let's start from the representation problem first.

Several peer traces are shown in figure 7. Peer trace can be drawn in 3D or 2D space as subfigure (a) and (b) respectively. In figure 7 (a) the vertical direction represent time, right is Ldeg and left is Udeg. In figure 7 (b) the vertical direction represents Udeg and horizontal Ldeg. 3D and 2D representations have their advantage and drawback respectively. For example look at the two subplots at left bottom and left top in 3D and 2D graphs. Both of them have an outlier point at left most. We can tell it is a state drift in the top and state setup at the bottom only in 3D graph. But in 2D graph, the state spreading area can be viewed more clearly. Traces in figure 10 can be classified into 2 main classes: drift within one mode (at USR as red dot subplots and LSR as blue dot subplots) and mode churn between USR and LSR (middle right subplot).

Along with the fast development of network and computers, today's Gnutella is not a network connected by small number of supper equipped machines and vast poor hosts. In our measurement, about 60% observed peers have been ultra at least once during the all observation time period. That means the mode of a peer is mainly determined by protocol and current conditions not its capabilities. The mode change is inevitable for most peers. In the section of queue modeling, we will point out that the mode change may be the most important factor to guarantee the ultra degree stability in Gnutella network.

**3.2 Peer Trace Classification:**

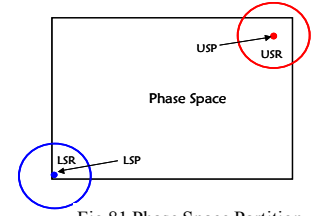

Fig 81 Phase Space Partition

There are thousands peer traces in out measurement. To view every trace by human eye is time consuming and unnecessary for understanding the general nature of these traces. Some classification method must be found to divide these traces into different groups with different trace evolution styles. The main evolution styles we concerned are the main stay region and state drift. Standard K-Means Clustering method is tried first since we think most traces will around LSR and USR two local centers. Then the method of LOF [15] is tried since the number of state drift is small at any specific trace but the event of state drift happens at almost all traces. We think it might be treated as outlier at every trace. Both above methods failed to classify traces drift around transition belt. Final classification method we adopted is simple but works well.

Our method first partition phase space $S$ into three regions like figure 8: ultra stable region (USR), leaf stable region (LSR) and transition region (TR). For USR and LSR, we first chose their center as ultra stable point (USP), leaf stable point (LSP) and radius $r_u$ and $r_l$ (both 10 at this paper). Then define

$USR = \{s \in S : d(s, USP) \leq r_u\}$

$LSR = \{s \in S : d(s, LSP) \leq r_l\}$

$TR = S \setminus (USR \cup LSR)$.

Table 1 Classification of LimeWire Traces
(Those with at least 10 responses)

| Peer Trace Attribute | | Trace Class | Sub plot | Peers |
|---|---|---|---|---|
| $\eta_l > 0, \eta_t = 0, \eta_u = 0$ | | Stable leaf | 1 | 771 (26.0%) |
| $\eta_l = 0, \eta_t > 0, \eta_u = 0$ | | Never stable ultra | 2 | 56 (1.9%) |
| $\eta_l = 0, \eta_t = 0, \eta_u > 0$ | | Stable ultra | 3 | 759 (25.6%) |
| $\eta_l > 0, \eta_t = 0, \eta_u > 0$ | | Bipolar between leaf and ultra | 4 | 215 (7.3%) |
| $\eta_l > 0, \eta_t > 0, \eta_u = 0$ | | Unstable Leaf | 5 | 50 (1.7%) |
| $\eta_l > 0, \eta_t > 0, \eta_u > 0$ | | Total churn | 6 | 378 (12.8%) |
| $\eta_l = 0$, $\eta_t > 0$, $\eta_u > 0$ | $\xi_t = 1$ | Stable Ultra (Occasional churn) | 7 | 468 (15.8%) |
| | $\xi_u = 1$ | Never stable ultra | 2 | 29 (0.98%) |
| | $\xi_t \leq \xi_u < 1$ | Half stable ultra | 8 | 74 (2.5%) |
| | $1 > \xi_t > \xi_u$ | Half unstable ultra | 9 | 161 (5.4%) |

We use a 6-tuple attribute $a_p=(\eta_l,\eta_t,\eta_u,\xi_l,\xi_t,\xi_u)$ to represent a peer trace, where $\eta_l$, $\eta_t$, $\eta_u$ are the number of reported states in that trace fall in LSR, TR and USR respectively normalized by the number of total reports, $\xi_l$, $\xi_t$, $\xi_u$ are the number of the trace enter or leaf the given region of LSR, TR and USR respectively normalized by the number of total reports in the given region. With this attribute, the classification on our measured LimeWire traces (with at least 10 responses) is listed in table 1. There are total 2961 traces. About half of them are stable. A quarter is stable leaf and a quarter is stable ultra in total. Another approximate quarter (little bit less) is stable in the sense that, they either jump in between of the leaf and ultra mode but stable whenever in each mode (about 7.3%), or mainly in USR with occasional drift (15.8%). Left traces (little bit more than a quarter) can be classified as unstable. They can be classified further as: Never stable ultra (about 3%), which working around the TR region never or rare enter USR, Unstable leaf (1.7%), which drift between LSR and TR, Total churn (12.8%), which travel all three regions, Half stable ultra (2.5%) and Half unstable ultra (5.4%), both of them walk around USR and TR, the only difference is the time they stay at USB and TR. In figure 9, we draw traces in different classes. In the 3D graph, the vertical direction is time and the time is accumulated in each subplot for showing all traces in one graph.

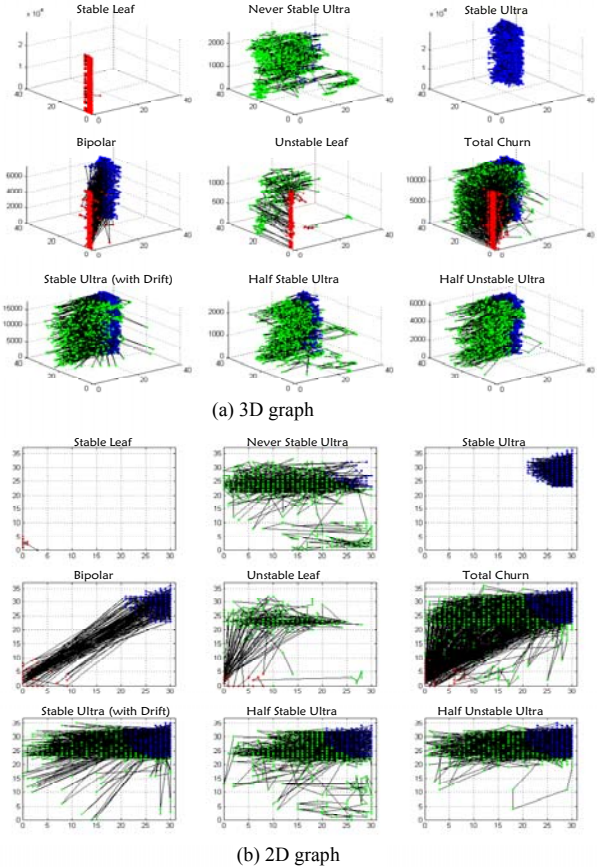

Fig 9 Traces of LimeWire in different classes

**3.3 Summary for this Section:** Peer traces are important bridge to relate the behavior of individual peer into the whole topology character of Gnutella network. Only very limited work has been done in this paper since it is new for us and we think for most researchers in this field as well. The summaries of this section as:

*Two Type of Mode churn: Core Connection Threshold* (the minimal connections of a ultra mode peer to other ultra peers) is the main feature for peer up to ultra from leaf mode or down to leaf mode from ultra. A new ultra peer can always capture specified number of other ultra peers in very short of time in guarantee the minimal connectivity of the core. An ultra peer will degrade into a leaf immediately if the minimal connectivity of the core cannot be maintained any more. *Kick Out* [18] can be observed occasionally (class of bipolar, subplot 4). In this case a redundant ultra will disconnect almost all of its current connections and become a leaf. Many interested questions such as "do most peers change mode during a long period?" "How long a peer will stay in leaf mode as well as in ultra mode?" have not been answered in this paper. We will try to answer them in further studies. Our peer traces in this paper might be too short for them we will try to enlarger our measurement duration.

*Ergodicy in The Drift on Ultra Mode:* Certain ergodicy do exist in the peer traces in sense the number of traces drifted in different areas is roughly proportional to the probability of these area induced from assemble degree distribution. But peer trace on ultra mode is not totally ergodic. Substantial peers drift in a smaller area around USP (class of stable ultra, subplot 3) and almost same peers drift in a much bigger areas (subplots 2, 7-9). These might be the "onion like topology" of [9] in a different point of view.

## IV. Degree and Connection Stability

**4.1 Degree Changes and Connection Changes:** Based on our measurement we believe that, the most important topological feature of Gnutella is its relative stable in connection degree and unstable in connection itself. In this section we will study this phenomenon.

Similar to peer degree trace we have discussed in previous section, the degree change and connection drop can also be represented as a trace. But the graph is too messy to be viewed clearly. Instead the center of traces, deviations of degree changes and connection changes can be drawn in the previously defined classifications. The degree change is defined as the norm of difference of consecutive reported degree. The connection change, we also call it connection departure or connection drop, is defined as the number of connections that is in previous report but not in the next report. The norm of change centers is defined. The bigger the norm is the larger the change of the trace. The degree change and connection change are small for stable leaf peers. It is due to the number of connections a leaf connects is mainly 2. The connection change is significantly larger than degree change in all other classes. For example in the class of stable ultra, the norm of degree change is around 2 but the connection change is around 7. That looks like the degree is controlled by a hidden process, the connection drop is a random disturbing and the room of dropped connections can be quickly filled by new connections. Different from the norm of change centers, the deviations of degree changes and connection changes are almost in same orders. It seems the hidden process could only compensate the number of connection drops but could not absorb the deviation of this change. We can draw following two interesting observations: first the changes in leaf connection and in ultra connections are correlated. Peer that loss more leaves will loss more ultras and vice versa. Second, the instability in mode is mainly due to large churn in ultra connections.

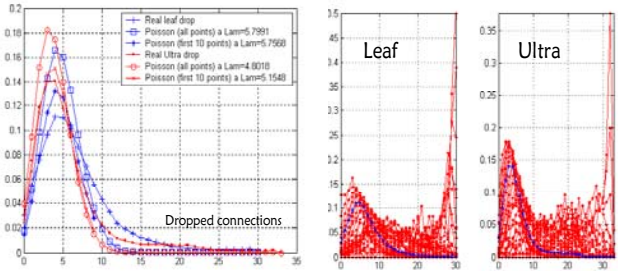

Fig 10 Connection Drop

**4.2 Connection Drop Rate and Live of Peers:** Figure 10 show the dropping probability of leaf and ultra connections for ultra mode (here we use Ldeg≥10 and Udeg≥10 as the criterion for ultra mode). The Poisson approximation is also drawn in this figure. It shows that the probability distribution for peer drop no matter leaf or ultra has a Poisson body and a slow dropping tail. The Poisson parameters we used here is simply the mean of the real distribution minus 1 to compassion the non-Poisson tail. In this case the maximum value of real distribution and Poisson distribution are best matched. The mean of Poisson distribution is about 5.8 for leaf and 4.8 for ultra. Since the crawl interval is half hour that result a departure rate about 11.6 leaf/hour and 9.6 ultra/hour. Only the first 11 points match is also depicted in the figure. This time we used the mean of those 11 points directly (without minus 1) as the parameters of Poisson. The mean is 5.0 and 4.3 for leaf and ultra respectively. The average connected leaf and ultra is 27.8507 and 29.9443 respectively. If the Poisson model can be used then the average departure rate of one single peer is about 0.42 and 0.32 for leaf and ultra connection respectively. That will result an average live of 2.4 and 3.1 hours for leaf and ultra connections. Next we will show that the live of we have crawled peers is much longer than the live we deducted from connection drop probability. It seems that the connection's live is much short than the live of peer which makes the connection. Since the dropping probability is resulted from the peers we have not crawled consistently, their real live we cannot know exactly. Another parallel hypothesis can also true that the peers we have not crawled have a live much short that that we have crawled. Since a connection is a bilateral busyness, our conclusion that connections live is much short than peers live might only partially true. The right two subplots in figure 10 show that, the distribution of leaf and ultra departure are not independent. The red lines in middle subplot are distributions of leaf departure conditioned on different ultra departures. The blue line is the marginal distribution of leaf departure. If we calculate the correlation coefficient, we can find the most correlation seems at the tail. The overall correlation coefficient is o.6128, but if only first 11 points are counted, the correlation coefficient dropped to 0.2531. It seems that, the small drop in connections are independent for leaf and ultra. But large amount of drop in leaf will accompany large drop in ultra and vice versa. The rightmost subplot is similar to the middle one; only interchange the position of "leaf" and "ultra".

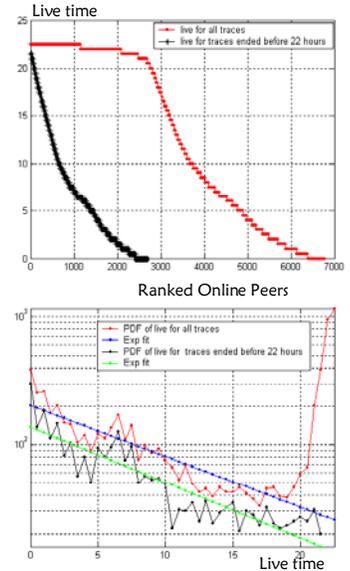

Fig 11 Live time Distribution

To find the live of peers is not so easy as first look at. The main reason is due to that not all peers make response at each crawl. So we cannot sure a miss responded peer is dead or is busy in other process and failed to response this crawl. Some break time *T* must be introduced to differentiate the dead and busy peers. If the time between two consecutive responses beyond the break time, we will count the peer is off line after the previous response and on line again at the successive response. Our calculation shows that there is no significant difference when the break time exceeds 2 hours. Hence in the following discussion we will set the break time equals 2 hours. Figure 11 shows the live of peers we have crawled. Since our crawl ended after 23 hours, we do not know the live of peers that have responded at the last two crawl periods (since the BearShare only response in a period of one hour). So there is a long flat line in the live time curve (the red line) of all peers in the top subplot. This will be disappeared if only the live of peers ended before 22 hours (the black line). The bottom subplot is the PDF of live time. The live time is roughly exponentially distributed with a mean of 0.092 for all peers and 0.10 for peers ended before 22 hours. That results a live of 10~11 hours, much longer than the connection live of 2~3 hours we have deducted before. The live of peers with different modes are depicted in figure 12. Since we do not know the mode of BearShare peers, the criterion that *Ldeg*≤2 and *Udeg*≤10 is used to determine whether a peer is in leaf or ultra mode if the mode is not reported directly. Thus there are 4021(59.33%) ultra only peers, 2017 (31.09 %) leaf only peers and 649(9.58%) churn mode peers. The live of stable peers is roughly exponentially distributed with a mean live time 11.23 hours for ultra only and 7.8hours for leaf only peers. The distribution of churn peers has a zeros slop. It can be approximately fitted by a uniform distribution. What to our surprise is that the live of leaf no matter peers or connections is short than ultra, but with no significant differences as we expected before. This further verified an observation we have made before that the mode of Gnutella is only a role network or/and protocol assigned to a user. It does not necessarily reflect the capability of the users host and network; neither reflect the intention of that user in using Gnutella network.

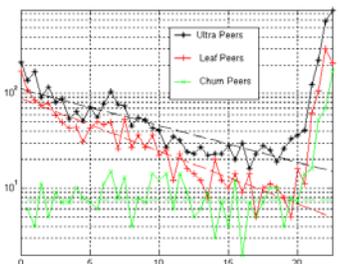

Fig 12 Live for different modes

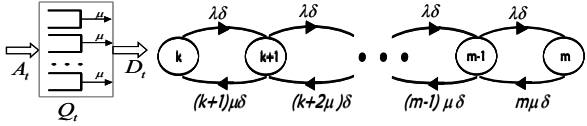

Fig 13 M/M/m/m System and Its State Transition

### V. MODELING THE DEGREE KEEPING PROCESS ON ULTRA MODE

**5.1 Degree keeping and M/M/m/m Queue:** Based on our analysis on previous section, the live time of connections is roughly exponentially distributed, the dropping rate of a connection is roughly Poisson, hence the Gnutella's degree keeping process on ultra mode can be approximately modeled as an M/M/m/m the m-server loss system [16,17]. For ultra mode peer, there are a maximum number of connections in Gnutella. If the connected number of peers reaches this number, a peer will never admit new connection. We will denote this number as $B_l$ for leaf connections and $B_u$ for ultra connections. For ultra mode peer, there is also a minimum number $L_u$. If the connected ultra peers are less than $L_u$, a peer will try to find and connect to other ultra peers by himself. If the connected ultra peers beyond $L_u$, a peer will passively wait other peers to initial the connection. Finding and connecting actively is much fast than waiting to be connected passively and the time to collect enough peers can be ignored based on our measurement. If we interpret the allowed connections as the servers, the connecting efforts as customers and the connection time as service time, the Gnutella degree keeping process is exactly an m servers loss system. As we discussed in previous section, the departure rate and service time in Gnutella is approximate memory less, hence an M/M/m/m will be a good model for this process. In figure 13, an M/M/m/m system and its state transition diagram is shown. Double M/M/m/m systems are needed in modeling Gnutella one for leaf connections and one for ultra connections, we will name them as LM and UM respectively. In LM, $m=B_l$ and $k=0$. In UM, $m=B_u$ and $k=L_u$. If the new connection effort rate is denoted as $\lambda_l$ and $\lambda_u$, and the connection dropping rate is denoted as $\mu_l$ and $\mu_u$ for leaf and ultra respectively, the generating matrix $g_l$ and $g_u$ for leaf and ultra respectively can be written as

$$g_l = \begin{bmatrix} -\lambda & \mu & 0 & ... & 0 \\ \lambda & -\lambda-\mu & 2\mu & ... & ... \\ 0 & \lambda & ... & 0 & 0 \\ ... & 0 & \lambda & -\lambda-(B_l-1)\mu & B_l\mu \\ 0 & 0 & 0 & \lambda & -B_l\mu \end{bmatrix}$$

$$g_u = \begin{bmatrix} -\lambda & (L_u+1)\mu & 0 & ... & 0 \\ \lambda & -\lambda-(L_u+1)\mu & (L_u+2)\mu & ... & ... \\ 0 & \lambda & ... & 0 & 0 \\ ... & 0 & \lambda & -\lambda-(B_u-1)\mu & B_u\mu \\ 0 & 0 & 0 & \lambda & -B_u\mu \end{bmatrix}$$

The equilibrium distributions $p_l^*$ and $p_u^*$ for leaf and ultra is respectively

$$\begin{cases} p_{l,i}^* = p_{l,0}^* (\lambda_l/\mu_l)^i / i! \\ p_{l,0}^* = \left[\sum_{N=0}^{B_l} (\lambda_l/\mu_l)^i / i!\right]^{-1} \end{cases}$$

$$\begin{cases} p_{u,i}^* = p_{u,0}^* (\lambda_u/\mu_u)^i L_u!/(L_u+i)! \\ p_{u,0}^* = \left[\sum_{n=0}^{B_u-L_u} (\lambda_u/\mu_u)^i L_u!/(L_u+i)!\right]^{-1} \end{cases}$$

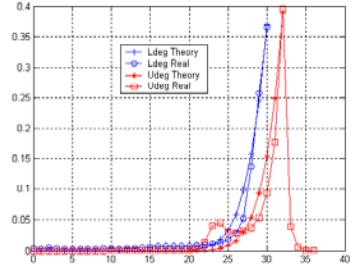

Fig 14 PDF of Degree

The transfer probability matrix for leaf and ultra is $T_l=exp(g_l)$ and $T_u=exp(g_u)$ with certain probability normalizations such that each column of these matrix are one. If we take the unit time as half an hour as our measurement, then the inferred departure rate is about $\mu_l=0.21$ and $\mu_u=0.16$.

Figure 14 depicts the real measured degree distribution for LimeWire and calculated degree based on M/M/m/m model. In this calculation we take $B_l=30$, $B_u=32$ and $L_u=20$, the designing parameters of LimeWire we inferred from our measurement results. The new connection effort $\lambda_l$ and $\lambda_u$ can be estimated from measurement results. Let $q$ be the probability of measured stable point, $u$ be the measured mean departure, we can use the formula $\lambda(1-q)=u$ or $\lambda=u/(1-q)$ to estimate the value of $\lambda$. For example, the measured $q$ for leaf and ultra is about 0.36 and 0.39, and the measured $u$ for leaf and ultra is about 5.8 and 4.8 respectively, will result the value of $\lambda$ to be 9.1 and 7.9 respectively. If the arriving rate below these number, the resulted degree distribution will lower than the real measured one that means more churn in degrees. The value used in figure 14 is $\mu_l=9.5$ and $\lambda_u=8$. It is worthwhile to notice that the loss probability of M/M/m/m is happening to be the probability of the stable point. That means that the success rate for each new connection enquiry is about 0.63 for leaf and 0.61 for ultra. Within half an hour, each peer will receive about 9.5 new leaf connection enquiries and among them about 5.98 are admitted and 3.5 are rejected. For ultra connections, about 4.89 are admitted and 3.1 are rejected. It seems need more connection efforts than the dropped connections if the dropped connections are go away permanently. The things much more like that, peer drops connections for other new connections. The stability in degree is ensured by unstable in connections.

**5.2 Degree Trace Generator For Ultra Peers:** One possible usage of above double M/M/m/m model is to generate degree sequence for simulation or simulation based studies. A random trace can be generated from given initial state.

For given transfer probability matrices $T_l$ and $T_u$ for leaf and ultra respectively, two new matrices $C_l$ and $C_u$ can be calculated such that each column of $C$ is the cumulated sum of $T$:

$$\begin{cases} C_l(i,j) = \sum_{k=0}^{i} T_l(k,j), \forall i,j \\ C_u(i,j) = \sum_{k=0}^{i} T_u(k,j), \forall i,j \end{cases}$$

A random sequence uniformly distributed in [0,1] then is generated as $\vec{r} = \{(r_{l,t}, r_{u,t})\}$. Form any given initial state $x_0=(x_{l,0}, x_{u,0})$, the trace can be calculated iteratively as:

$$\begin{cases} x_{l,t+1} = k_l, \text{ if } C_l(k_l-1, x_{l,t}) \le r_{l,t} < C_l(k_l, x_{l,t}) \\ x_{u,t+1} = k_u + L_u, \text{ if } C_u(k_u-1, x_{u,t}-L_u) \le r_{u,t} < C_u(k_u, x_{u,t}-L_u) \end{cases}$$

Above discussed M/M/m/m system is based on infinite small time interpretation. So we will call it Continue Time Double M/M/m/m System (CTDM). The advantage of CTDM is easy to find analytical solutions to this system. The drawback of it is that the departure process is hidden in this presentation. Some time we need to represent the departure process explicitly since this process can be measured directly, so we wish to seek other way to describe the double M/M/m/m system.

A direct method is to use binomial distribution to describe the departure process. For example the leaf degree at time $t$ is $q_{l,t}$, then departure probability can specified by a probability generating function

$$g(x, q_{l,t}) = (\mu_l x + (1-\mu_l))^{q_{l,t}}.$$

Then the transfer matrices $T_l$ and $T_u$ for leaf and ultra Markov chains can be written as:

$$T_{i,j}^{(l)} = \begin{cases} \sum_{k=\max(0,j-i)}^{j} \binom{j}{k} \mu_l^k \bar{\mu}_l^{j-k} \frac{\lambda_l^{i-j+k}}{(i-j+k)!} e^{-\lambda}, i < B_l \\ \sum_{k=0}^{j} \binom{j}{k} \mu_l^k \bar{\mu}_l^{j-k} \sum_{i=B_l}^{\infty} \frac{\lambda_l^{i-j+k}}{(i-j+k)!} e^{-\lambda}, i = B_l \end{cases}$$

$$T_{i-L_u, j-L_u}^{(u)} = \begin{cases} \sum_{k=\max(0,j-i)}^{j} \binom{j}{k} \mu_u^k \bar{\mu}_u^{j-k} \frac{\lambda_u^{i-j+k}}{(i-j+k)!} e^{-\lambda}, i < B_u \\ \sum_{k=0}^{j} \binom{j}{k} \mu_u^k \bar{\mu}_u^{j-k} \sum_{i=B_u-L_u}^{\infty} \frac{\lambda_u^{i-j+k}}{(i-j+k)!} e^{-\lambda}, i = B_u \end{cases}$$

The equilibrium distributions $p_l^*$ and $p_u^*$ of leaf degree and ultra degree can be solved from these matrices. We will call this system as Binomial Discrete Time M/M/m/m System (BDTM).

Use BDTM to generate degree trace is relatively simple. In each iterate step, we generate two Poisson distributed random number $A_l$ and $A_u$ according to arrive rate $\lambda_l$ and $\lambda_u$ respectively. And for given degree $d_{l,t}$ and $d_{u,t}$, we generate two [0, 1] uniformly distributed random sequences $R_l = \{r_{l,1}, \cdots r_{l,d_{l,t}}\}$ and $R_u = \{r_{u,1}, \cdots r_{u,d_{u,t}}\}$, and counter the number $n_l = |\{(r_{l,i} \in R_l, r_{l,i} \le \mu_l)\}|$ and $n_u = |\{(r_{u,i} \in R_u, r_{u,i} \le \mu_u)\}|$ as the number of dropped connections. The next degree can be calculated as

$$\begin{cases} d_{l,t+1} = d_{l,t} + A_l - n_l, \\ d_{u,t+1} = d_{u,t} + A_u - n_u \end{cases}$$

Set the degree equal to the nearest boundary, if the calculated degree outside the legal region.

The problem to use BDTM model is how to chose the time scale. If the same parameters as CTDM is used, the resulted degree distribution will be significant different from that of CTDM. For same degree distributions, the parameters must be scaled.

**VI. Conclusions**

In this paper, the topology dynamics of Gnutella are studied based on measurement, theoretical modeling and simulations. Phase space is introduced in this paper to interpret and analysis of the degree evolution of Gnutella peers. The dynamic behavior of peer's connection can be represented and interpreted more clearly. The phase space can be partitioned into different regions some how related to peer's software version. The topological status of a peer is tightly related to the region it falls in the phase space. With this space partition, a classification method is introduced to classify observed traces into different classes that reflect peers overall topological status among our measurement. Connection churn then is studied along with the churn in degree. Our data showed that the connections churn much more violent than that of degree. It seems that the topological structure of Gnutella is rather stable in its connection degree but not the topology itself. The connection drop rate is estimated and the live time of connections is inferred afterwards. The drop probability has a Poisson body and a heavy tail and the live time is roughly exponential distributed. The live time of a connection is about 2-3 hour much short than the live time of peers that last 8-11 hours in average. M/M/m/m loss queue system is introduced to model the degree keeping process in Gnutella and to generate random degree sequence in simulations. This model revealed that the degree stable is ensured by large new connection efforts. In other words the stable in topological structure of Gnutella is a results of essential unstable in its topology. That opens a challenge to the basic design philosophy of this network.